\documentstyle[eqsecnum,epsfig,aps,prd]{revtex}
\tighten
\let\jnfont=\rm
\def\NPB#1,{{\jnfont Nucl.\ Phys.\ B }{\bf #1},}
\def\PLB#1,{{\jnfont Phys.\ Lett.\ B }{\bf #1},}
\def\EPJC#1,{{\jnfont Eur.\ Phys.\ J.\ C }{\bf #1},}
\def\PRD#1,{{\jnfont Phys.\ Rev.\ D }{\bf #1},}
\def\PRL#1,{{\jnfont Phys.\ Rev.\ Lett.\ }{\bf #1},}
\def\MPLA#1,{{\jnfont Mod.\ Phys.\ Lett.\ A }{\bf #1},}
\def\JPG#1,{{\jnfont J.\ Phys.\ G}{\bf #1},}
\def\CTP#1,{{\jnfont Commun.\ Theor.\ Phys.\ }{\bf #1},}

\def\gsim{\mathrel{\mathpalette\oversim>}}
\def\oversim#1#2{\lower0.5ex\vbox{\baselineskip0pt\lineskip0pt
  \lineskiplimit0pt\everycr{}\tabskip0pt
  \halign{$\mathsurround0pt #1\hfil##\hfil$\crcr #2\crcr\sim\crcr}}}

\begin{document}
\draft
\preprint{}

\title{Dark Matter Constraints on Gaugino/Higgsino Masses in Split Supersymmetry\\ 
         and Their Implications at Colliders}                                                                                
                                                                                
\author{Fei Wang $^1$,  Wenyu Wang $^1$, Jin Min Yang $^{2,1}$}
                                                                                
\address{ \ \\[2mm]
{\it $^1$ Institute of Theoretical Physics, Academia Sinica, Beijing 100080, China} \\ [2mm]
{\it $^2$ CCAST (World Laboratory), P. O. Box 8730, Beijing 100080, China} \ \\[6mm] }
 
\date{\today}
 
\maketitle

\begin{abstract}
In split supersymmetry, gauginos and Higgsinos are the only supersymmetric 
particles which are  possibly accessible at foreseeable colliders.  
While the direct experimental searches, such as LEP and Tevatron experiments, gave 
robust lower bounds on the masses of these particles, the cosmic dark matter can 
give some upper bounds and thus have important implications for the searches at 
future colliders.
In this work we scrutinize such dark matter constraints and show the allowed mass 
range for charginos and neutralinos (the mass eigenstates of gauginos and Higgsinos). 
We find that the lightest chargino must be lighter than about 1 TeV 
under the popular assumption $M_1=M_2/2$ and about 2 or 3 TeV in other cases. 
The corresponding production rates of the lightest chargino at the CERN Large Hadron 
Collider (LHC) and the International Linear Colldier (ILC) are also shown. 
While in some parts of the allowed region the chargino pair production rate can be larger 
than 1 pb at LHC and 100 fb at the ILC, other parts of the region correspond to very small
production rates and thus there is no guarantee to find the charginos of split supersymmetry
at future colliders.  

\end{abstract}


\section{Introduction}
Given the importance of supersymmetry in both particle physics and string theory,
searching for supersymmetry seems to be a crucial task in the ongoing and forthcoming colliders.
The forthcoming LHC collider would be able to explore the supersymmetric particles 
up to a few TeV and the ILC collider would allow precision
test of  supersymmetry. Of course, there is no guarantee to find  supersymmetry at these collders 
since the masses of the supersymmetric particles are basically unknown. As is well known, 
in order to solve the fine-tuning problem in particle theory, the supersymmetric particles  should 
be below TeV scale and thus the LHC would be a factory of supersymmetric particles.  
However, in the recently proposed split supersymmetry \cite{split}, the supersymmetric solution of 
fine-tuning problem in particle physics is given up (inspired by the need of fine-tuning for the 
cosmological constant), while the virtues of supersymmetry in preserving grand unification as well 
as providing the cosmic dark matter candidate are still retained.  As a result, the mass scale
of all sfermions as well as the several heavy Higgs bosons  
can be very high while the gaugino/Higgsino mass scale may be still below the
TeV scale. While the split supersymmetry has the obvious virtue of naturally avoiding the 
notorious supersymmetric flavor problem,  it predicts that no supersymmetric scalar particles 
except a light Higgs boson are accessible at the foreseeable particle colliders. 
Thus, if split supersymmetry is the true story 
\footnote{Some studies \cite{split-natural}
showed that split supersymmetry is quite natural from the top-down view.}, 
the only way to reveal supersymmetry at the colliders is through gaugino or Higgsino productions.

Among the gauginos and Higgsinos, the gluino is the only colored particle and thus may be
most copiously produced in the gluon-rich environment of the LHC. 
However, the gluino is usually speculated to be much heavier 
than other gauginos and Higgsinos.  It was shown \cite{senatore} that the grand unification 
requirement can allow a gluino as heavy as 18 TeV. Furthermore, if the dark matter is assumed 
to be the gravitino produced from the late decay of the meta-stable gluino which freezed out 
at the early universe, it was found \cite{wang} that the gluino must be heavier than about 
14 TeV and thus impossibly accessible at the LHC. So, to explore split supersymmetry we should not 
focus only on gluino productions; the productions of the electroweakly interacting gauginos 
and Higgsinos (which mix into two charged particles called charginos and four neutral particles 
called neutralinos) should also be considered although their production rates at the LHC 
are much lower.
Of course, if the LHC can discover supersymmetry and then the ILC takes the task of precision 
test, the productions of charginos and neutralinos at the ILC will play the dominant role.
For both the LHC and the ILC, the production of charginos will give good signatures 
since the subsequent decays yield energetic leptons. To facilitate the collider searches for
the charginos and neutralinos, the pre-estimation of their allowed mass regions 
are important.  

While the direct experimental searches, such as LEP and Tevatron experiments, gave 
robust lower bounds for the masses of charginos and neutralinos, the cosmic dark matter 
can give some upper bounds and thus have important implications for the searches at future 
colliders. Therefore, although the consequence of split supersymmetry in the dark matter issue 
has been considered to some extent in the literature \cite{senatore,wang,pierce,profumo}, 
we in this work scrutinize the dark matter constraints on the masses of charginos and neutralinos 
and evaluate the corresponding production rates at the LHC and ILC. 

This work is organized in the follows. In Sec. II we recapitulate the parameter space of
the sector of charginos and neutralinos. In Sec. III we examine the dark matter constraints.
We will show the constraints on (a) the original parameter space, (b) the masses of charginos and
neutralinos, and (c) the production rates at the LHC and the ILC. 
The conclusions are given in Sec. IV.   

Note that for the supersymmetry parameters we adopt the notation in Ref.~\cite{gunion}.
We work in the framework of the Minimal Supersymmetric Model 
(MSSM) and assume the lightest supersymmetric particle is the lightest neutralino, 
which solely makes up the cosmic dark matter.  Also, we fix the parameter $\tan\beta=40$ since 
a large value of  $\tan\beta$ is favored by current experiments and, in the region of large
$\tan\beta (\gsim 10)$, our results are not sensitive to $\tan\beta$. 

\section{Parameter space of charginos and neutralinos} 
The chargino mass matrix is given by 
\begin{eqnarray} \label{mass1}
\left( \begin{array}{cc} M_2 & \sqrt 2 m_W \sin\beta \\
                        \sqrt 2 m_W \cos\beta & \mu \end{array} \right) ,
\end{eqnarray}
and the neutralino mass matrix is given by
\begin{eqnarray} \label{mass2}
\left( \begin{array}{cccc} 
M_1 & 0   & - m_Z \sin\theta_W \cos\beta & m_Z \sin\theta_W \sin\beta  \\
0   & M_2 & m_Z \cos\theta_W \cos\beta   &-m_Z \cos\theta_W \sin\beta  \\
- m_Z \sin\theta_W \cos\beta &  m_Z \cos\theta_W \cos\beta  & 0 & -\mu \\
m_Z \sin\theta_W \sin\beta & -m_Z \cos\theta_W \sin\beta & -\mu & 0\\
\end{array} \right),
\end{eqnarray}
where $M_1$ and $M_2$ are  respectively the $U(1)$ and $SU(2)$ gaugino mass parameters,
$\mu$ is the mass parameter in the mixing term $-\mu \epsilon_{ij} H_1^iH_2^j$ in the
superpotential, and  $\tan\beta \equiv v_2/v_1$ is ratio of the vacuum expectation values 
of the two Higgs doublets. The diagonalization of (\ref{mass1}) gives two 
charginos $\chi^+_{1,2}$ with the convention $M_{\chi^+_1}<M_{\chi^+_2}$;
while the diagonalization of (\ref{mass2}) gives four neutralinos $\chi^0_{1,2,3,4}$ 
with the convention $M_{\chi^0_1}<M_{\chi^0_2}<M_{\chi^0_3}<M_{\chi^0_4}$.
So the masses and mixings of charginos and neutralinos
are determined by four parameters: $M_1$, $M_2$, $\mu$ and $\tan\beta$.

The current constraints \cite{pdg} on these parameters are divided into two classes:
(1) direct constraints from experimental searches of supersymmetric particles; (2) 
indirect constraints from some precisely measured low-energy processes or
physical quantities via supersymmetric quantum effects. 
For split supersymmetry, almost all indirect constraints from low-energy processes,
such as various $B$-decays, drop out since the supersymmetric loop effects in 
these processes usually involve sfermions which are superheavy in split supersymmetry.
The most stringent direct bounds are from LEP experiments \cite{lep2-web}: 
(i) the lighter chargino $\chi^+_1$ must be heavier than about 103 GeV;
(ii) the LSP must be heavier than about 47 GeV;
(iii) the value of $\tan\beta$ must be larger than 2.

Note that in addition to the direct lower bound from LEP II, theoretically a large $\tan\beta$ 
helps to push up the lightest Higgs boson mass and thus ameliorate the stress 
between the experimental lower bound and the theoretical upper bound on 
the lightest Higgs boson mass \footnote{The upper mass limit of the lightest Higgs boson is
relaxed to about 150 GeV in split supersymmetry \cite{split}. However, if the right-handed 
neutrinos are introduced into split supersymmetry with see-saw mechanism, the large neutrino 
Yukaka couplings can lower the lightest Higgs boson mass by a few tens of GeV \cite{cao}.}. 
So in our analyses we assume a large $\tan\beta$
and fix it to be 40. We checked that in the region of large
$\tan\beta (\gsim 10)$, our results are not sensitive to $\tan\beta$. 

We assume the cosmic dark matter is solely composed of the LSP, which is
assumed to be the lightest neutralino $\chi^0_1$. (If the gravitino
is assumed to be the LSP and make up the dark matter, it will incur
some severe cosmoligical constraints \cite{feng}.)  

The thermal relic density of the lightest neutralino from the freeze-out
can be calculated from the Boltzmann equation which involves the thermal 
averaged cross section of neutralino annihilations. In our work we use the 
package DarkSUSY \cite{darksusy} and we checked that the package micrOMEGAs
\cite{micro} gives the similar results in our study. 
In calculating the
cross section of neutralino annihilations, many additional supersymmetric parameters 
are involved, among which the most important ones are sfermion masses and
$M_A$ (the mass of CP-odd Higgs boson). Since we focus only on split supersymmetry,
all sfermion masses and $M_A$ are superheavy. So any diagrams involving 
a sfermion or a heavy Higgs boson makes negligible contributions to neutralino 
annihilations. Actually, we found that as long as the sfermion mass or $M_A$      
gets heavier than about 10 TeV, the effects of sfermions or heavy Higgs bosons
decouple, as shown in Figs. 1 and 2. The peak in Fig.1(b) happens around the
'$A$-funnel' resonance point $M_A\approx 2 M_{\chi^0_1}$.  
\begin{figure}[tb]
\begin{center} \epsfig{file=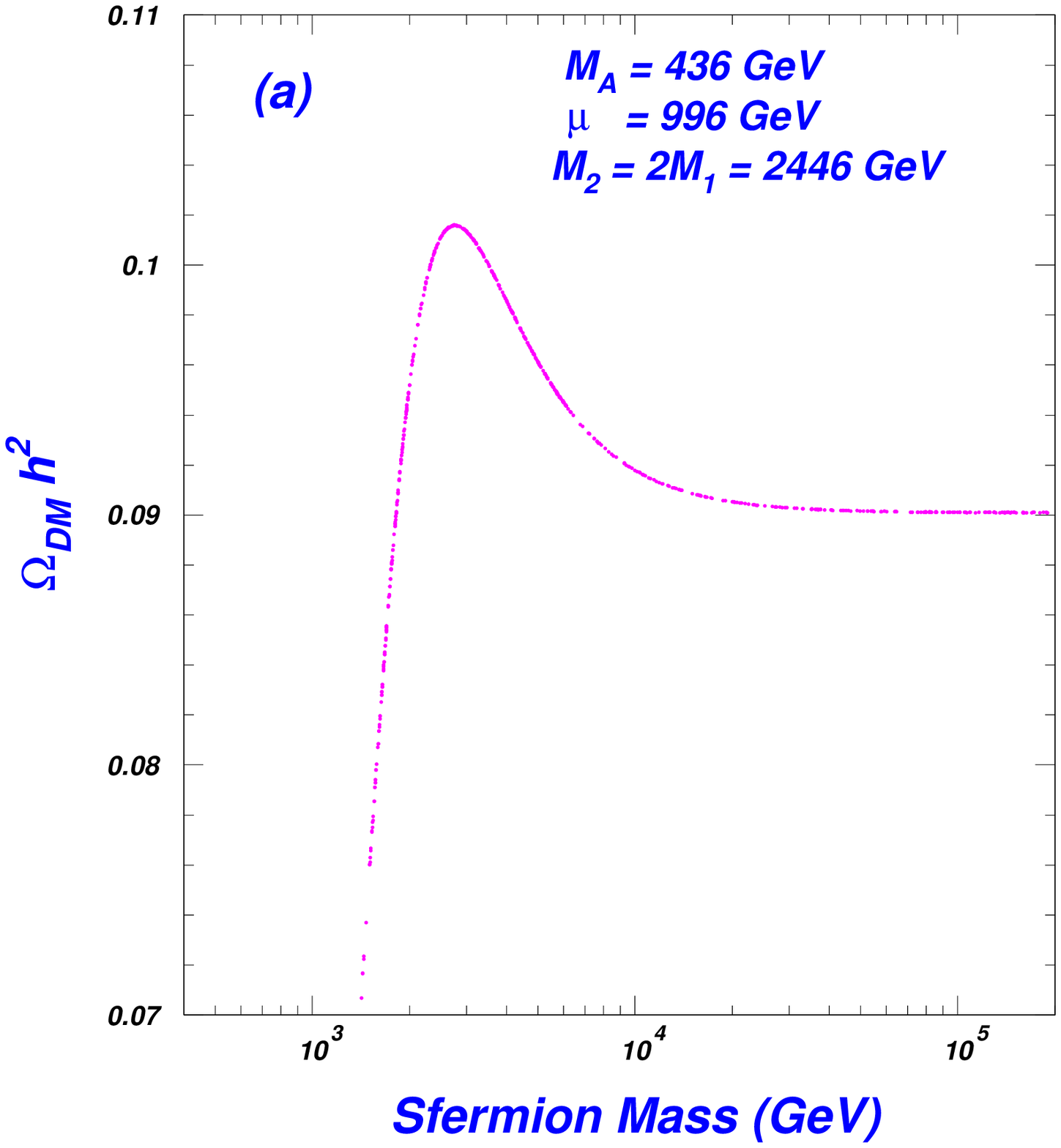,width=8.0cm, height=9cm}
               \epsfig{file=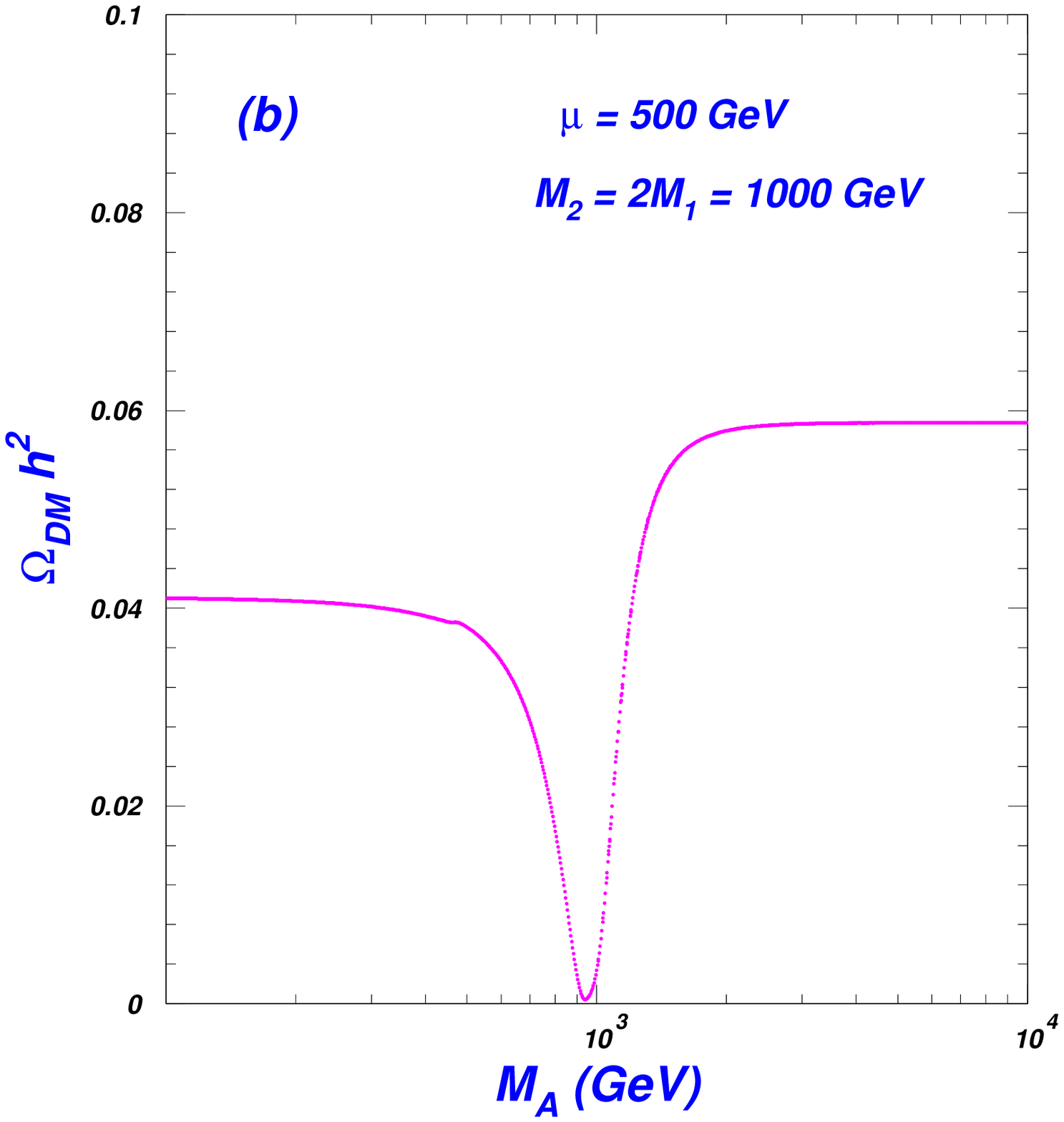,width=8.0cm, height=9cm}
\end{center}
\caption{The neutralino relic density versus (a) sfermon mass; (b) $M_A$. }
\end{figure}

\section{ WMAP dark matter constraints}
The $2\sigma$ allowed region for the dark matter relic density is 
\begin{eqnarray}
       0.094<\Omega_{CDM}h^2<0.129 ,
\end{eqnarray}
which can be inferred from the Wilkinson Microwave Anisotropy Probe (WMAP) 
measurements \cite{wmap}. 
In the following we present the $2\sigma$ allowed regions for four cases:
(1) $M_1=M_2/2$; (2) $\mu$ is superheavy; (3) $M_2$ is superheavy; and 
(4) $M_1$ is superheavy. 
In our calculations we fix a 'superheavy' mass to be 100 TeV since it is
high enough for the relevant supersymmetric particles to decouple from
the neutralino annihilations.
In each case we present the allowed region for   
(a) the original parameter space;
(b) the chargino mass $M_{\chi^+_1}$ versus the neutralino mass $M_{\chi^0_1}$;
(c) the cross section of chargino pair production at LHC and ILC versus 
the chargino mass $M_{\chi^+_1}$. Note that we just give the tree-level cross sections
and do not include the one-loop corrections \cite{char-lhc}.  
The center-of-mass energy is 14 TeV for LHC and is assumed to be 1 TeV for ILC \cite{ILC}. 
The results are resented in Figs. 2-5, where the dot-dashed line is the LEP II 
lower limit on chargino mass.   

{\em (1) $M_1=M_2/2$:} 
This case is well motivated since the supergravity models predict the unification 
relation $M_1=\frac{5}{3}M_2\tan^2\theta_W\simeq 0.5 M_2$. 
In the low mass region for both $M_2$ and $\mu$ in Fig. 2(a), the dark matter is 
the mixing of gauginos and Higgsinos. The strip with very large $M_2$ corresponds
to Higgsino dark matter, while the strip with very large $\mu$  corresponds to gaugino
dark matter (the mixing of bino and wino).     
From Fig. 2(b) we see that both the chargino mass $M_{\chi^+_1}$ and the neutralino 
mass $M_{\chi^0_1}$ is upper bounded by about 1 TeV. 
Fig. 2(c) shows that in the allowed region the cross section of chargino pair production
at LHC can reach a few pb for a light chargino, but drops rapidly as the chargino gets heavy. 
The cross section at ILC can reach 100 fb for a light chargino in the allowed region.  

{\em (2) Superheavy $\mu$:}
This case was proposed and favored by some authors \cite{split-split} because 
the $\mu$ problem \cite{mu} is avoided and a crude gauge coupling unification is preserved.
In the low mass region for both $M_2$ and $M_1$ in Fig. 3(a), the dark matter is 
the mixing of bino and wino; while the region with large $M_1$  corresponds to wino dark matter.
Fig. 3(b) shows that both the chargino mass $M_{\chi^+_1}$ and the neutralino 
mass $M_{\chi^0_1}$ is upper bounded by about 3 TeV. 
Fig. 3(c) shows that in the allowed region the cross section of chargino pair production
at LHC can reach  10 pb for a light chargino, but drops rapidly as the chargino gets heavy. 
The cross section at ILC can reach 200 fb for a light chargino in the allowed region.  
\begin{figure}[tb]
 \hspace*{-1.5cm} \epsfig{file=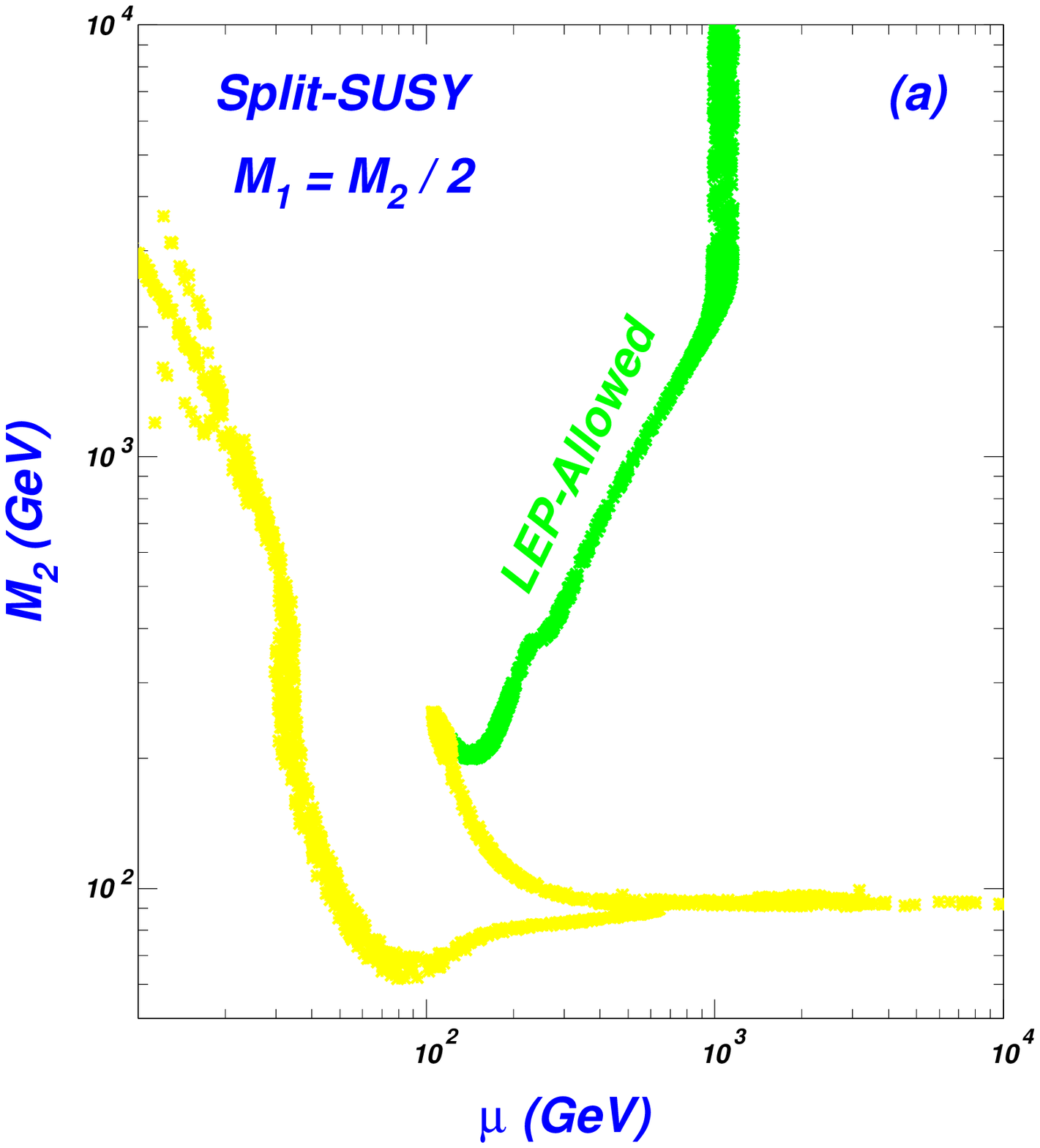,width=6.2cm, height=7.3cm}
                  \epsfig{file=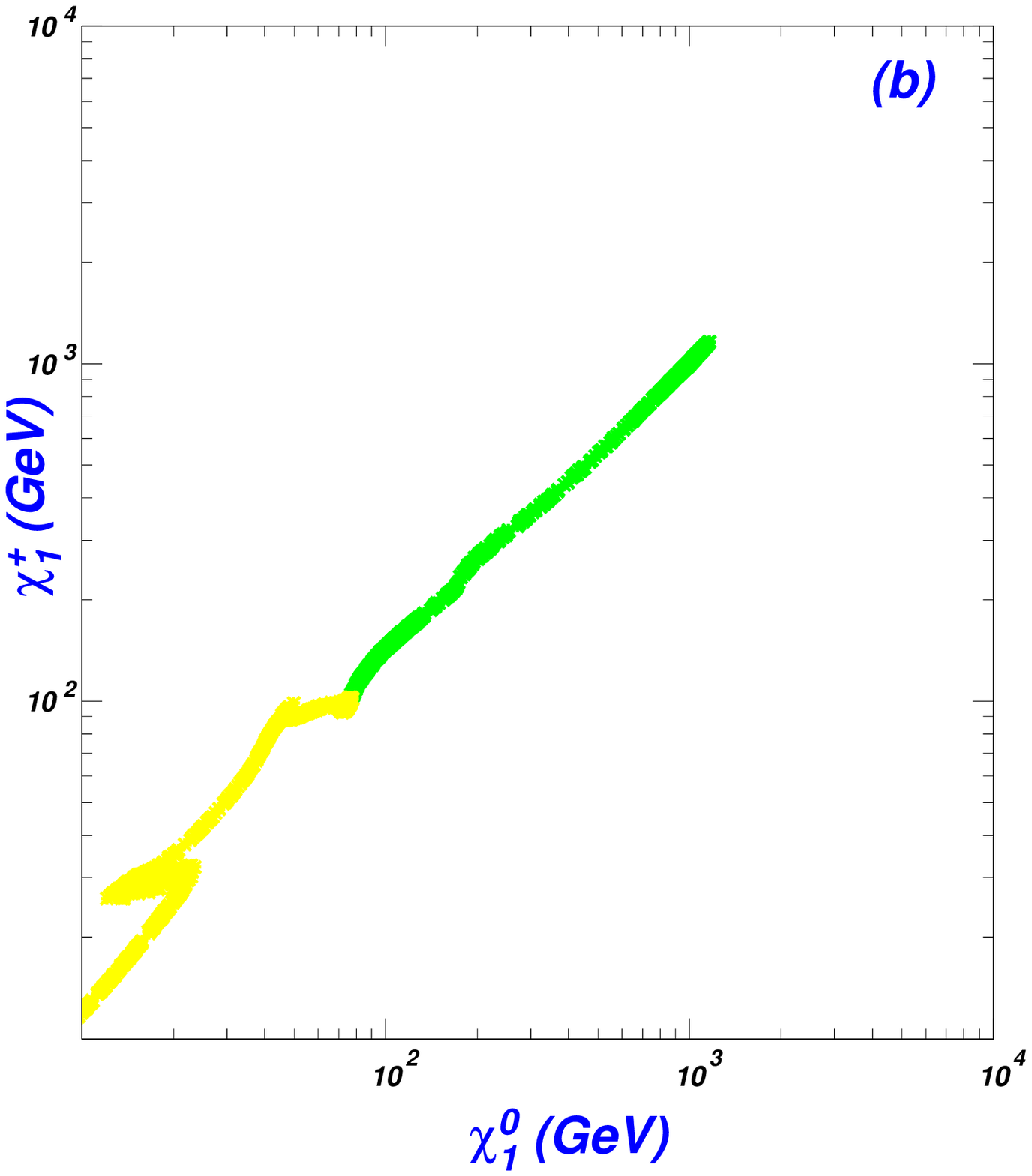,width=6.2cm, height=7.3cm}
                  \epsfig{file=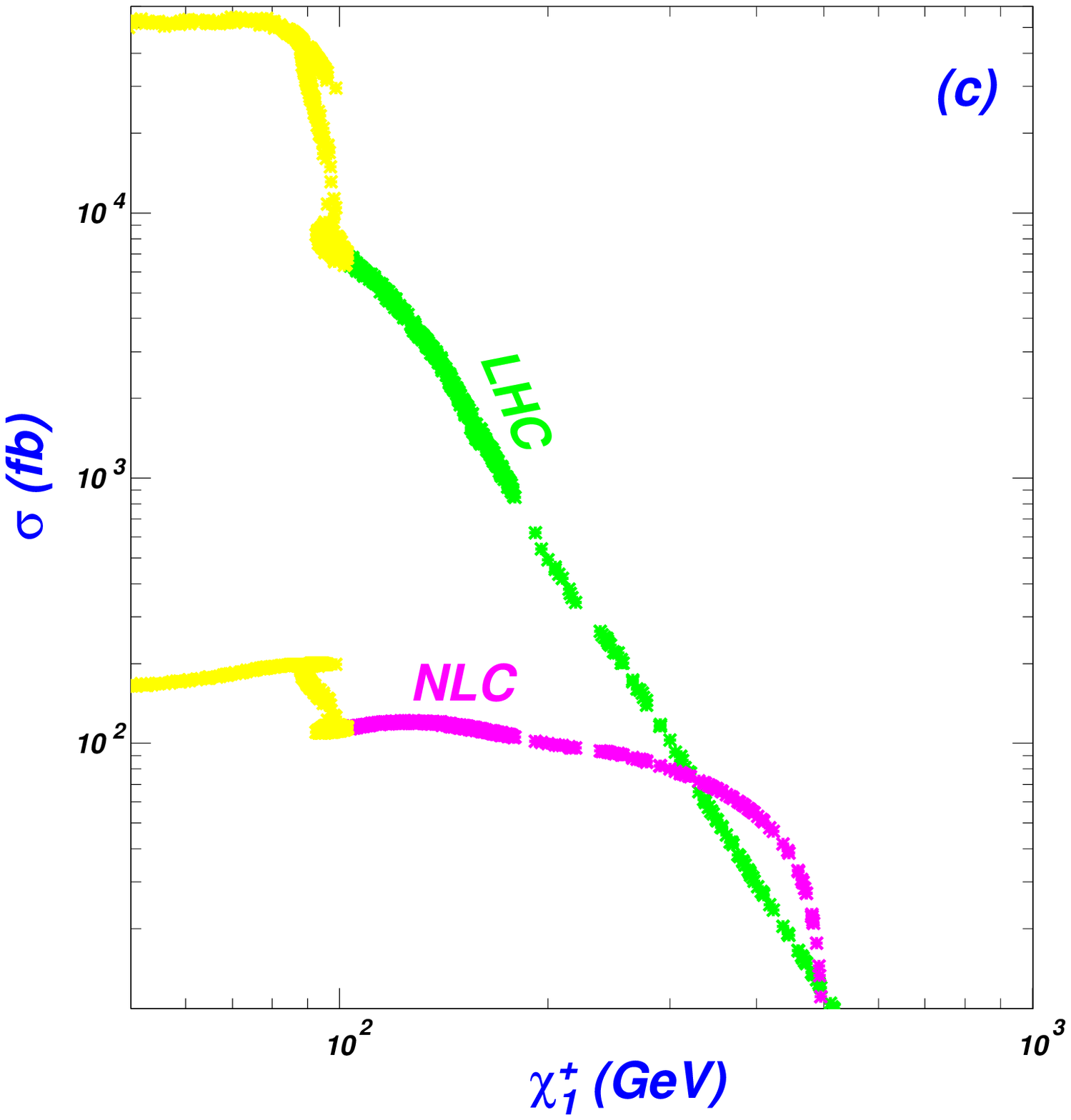,width=6.2cm, height=7.3cm}
 \vspace*{0.5cm}
\caption{The WMAP $2\sigma$ allowed region (shaded area) in case of $M_1=M_2/2$. 
         The light shaded region (yellow) is not allowed by LEP experiment.}
\end{figure}
\begin{figure}[tb]
 \hspace*{-1.5cm} \epsfig{file=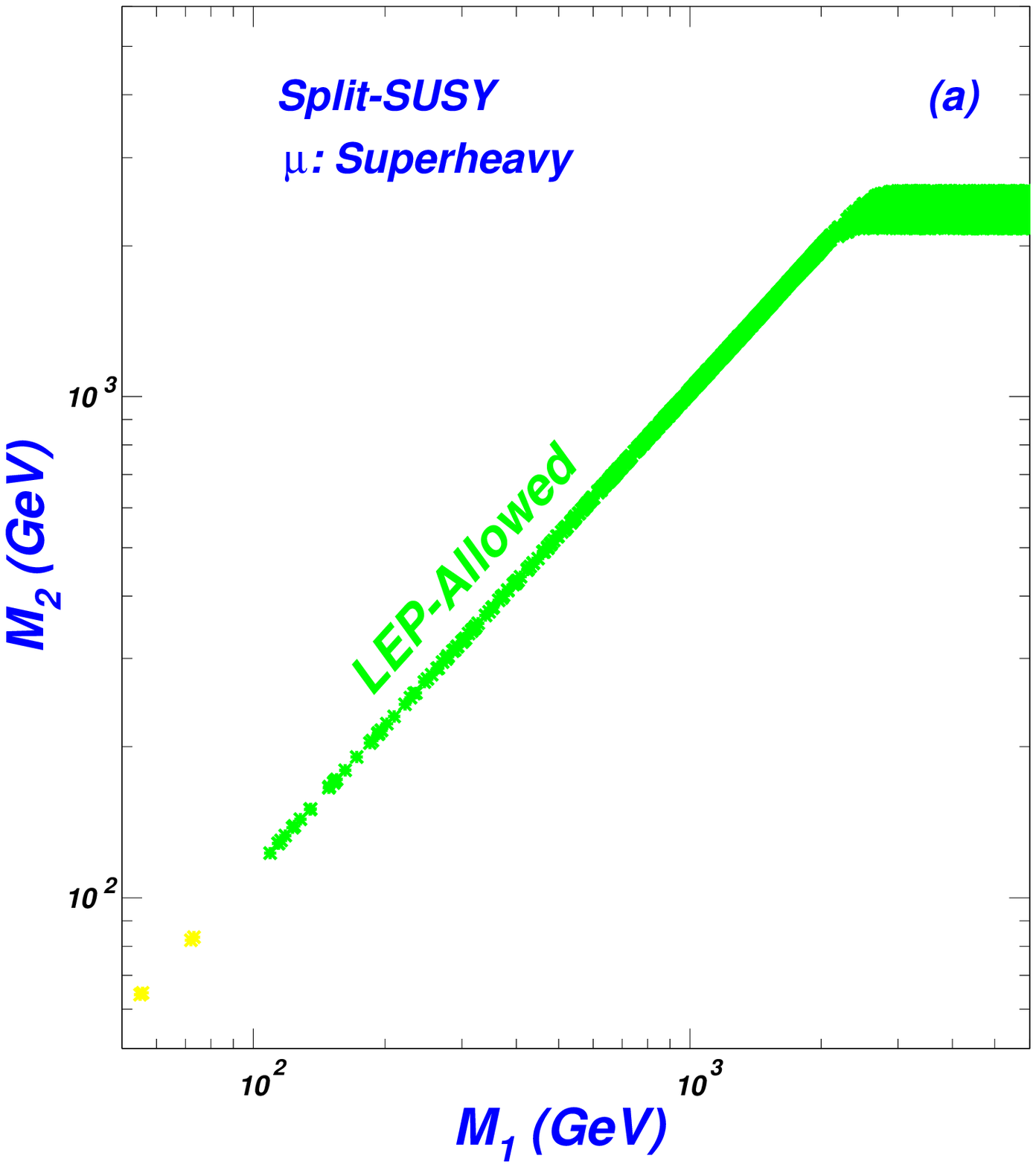,width=6.2cm, height=7.3cm}
                  \epsfig{file=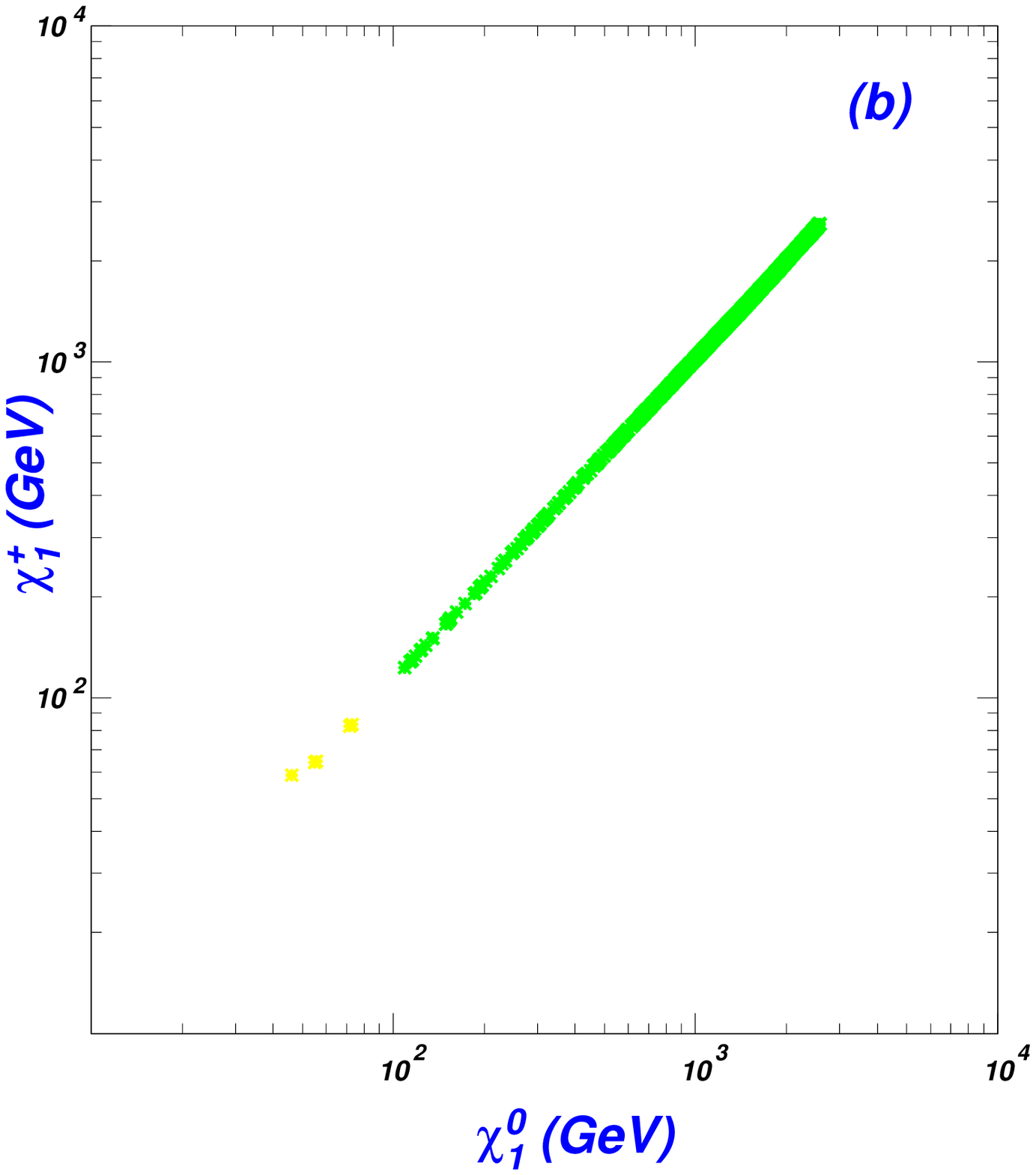,width=6.2cm, height=7.3cm}
                  \epsfig{file=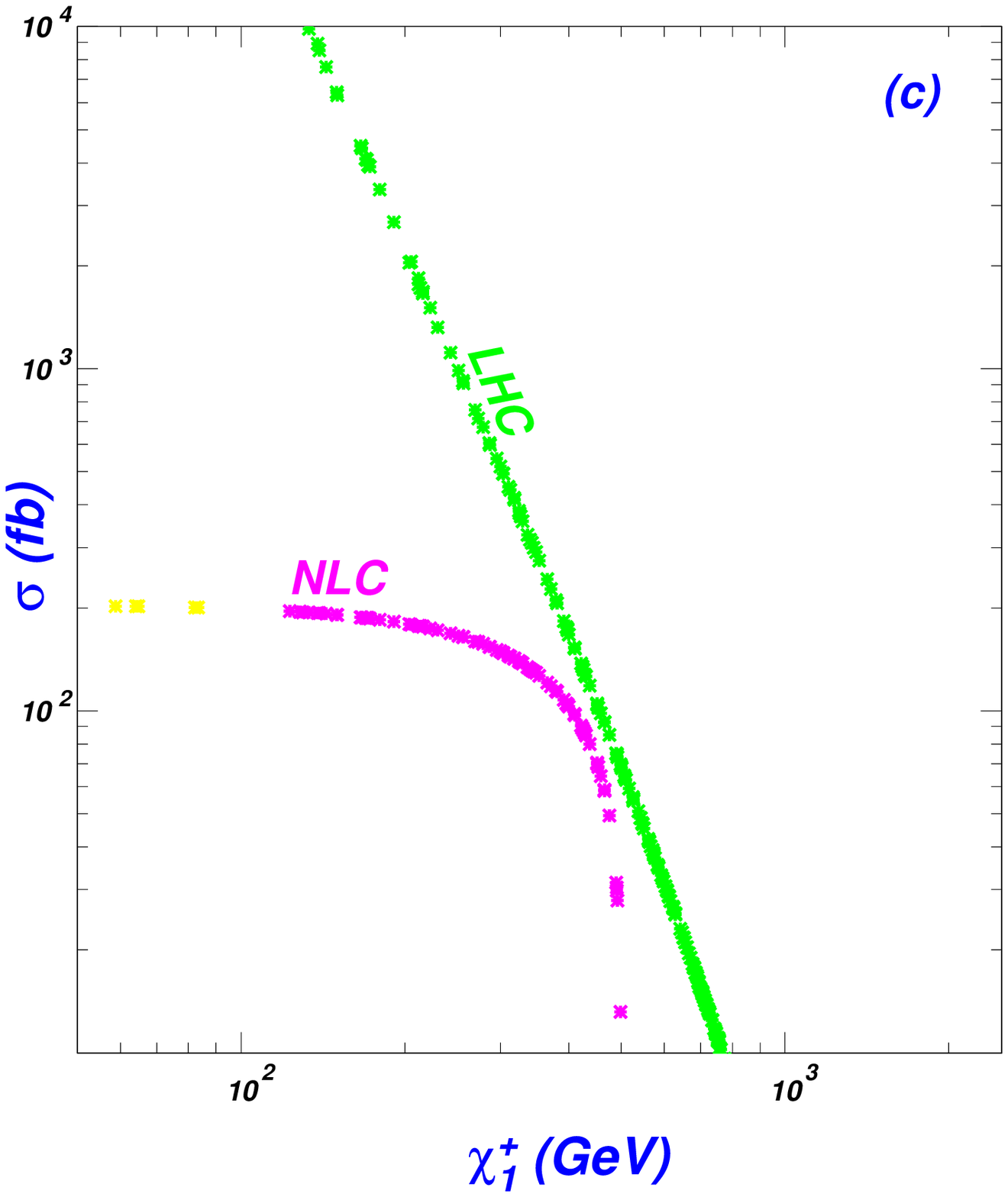,width=6.2cm, height=7.3cm}
 \vspace*{0.5cm}
\caption{The WMAP $2\sigma$ allowed region (shaded area) in case of superheavy $\mu$.
         The light shaded region (yellow) is not allowed by LEP experiment.}
\end{figure}

{\em (3) Superheavy $M_2$:}
In the low mass region for both $M_1$ and $\mu$ in Fig. 4(a), the dark matter is 
the mixing of bino and Higgsinos. When  $M_1$ ( $\mu$ ) gets very large, a strip is remained,
which  corresponds to Higgsino (bino) dark matter.  The chargino mass $M_{\chi^+_1}$ 
is upper bounded by about 3 TeV and the neutralino mass $M_{\chi^0_1}$ is upper bounded by 
about 1 TeV. For a light chargino in the allowed region, the cross section of chargino pair 
production can reach the level of pb at LHC and 100 fb at ILC.

{\em (4) Superheavy $M_1$:}
In Fig. 5(a) the strip with large  $M_2$ ($\mu$) corresponds to Higgsino (wino) dark matter.  
As shown in Fig. 5(b), both the chargino mass $M_{\chi^+_1}$ and the neutralino mass 
$M_{\chi^0_1}$ are lower bounded by about 1 TeV and upper bounded by about 2.5 TeV.
Thus the charginos cannot be pair produced at the ILC with c.m. energy of 1 TeV.  
Although the charginos can be  pair produced at the LHC, the cross section is very small,
as shown in Fig. 5(c). 

\begin{figure}[tb]
 \hspace*{-1.5cm} \epsfig{file=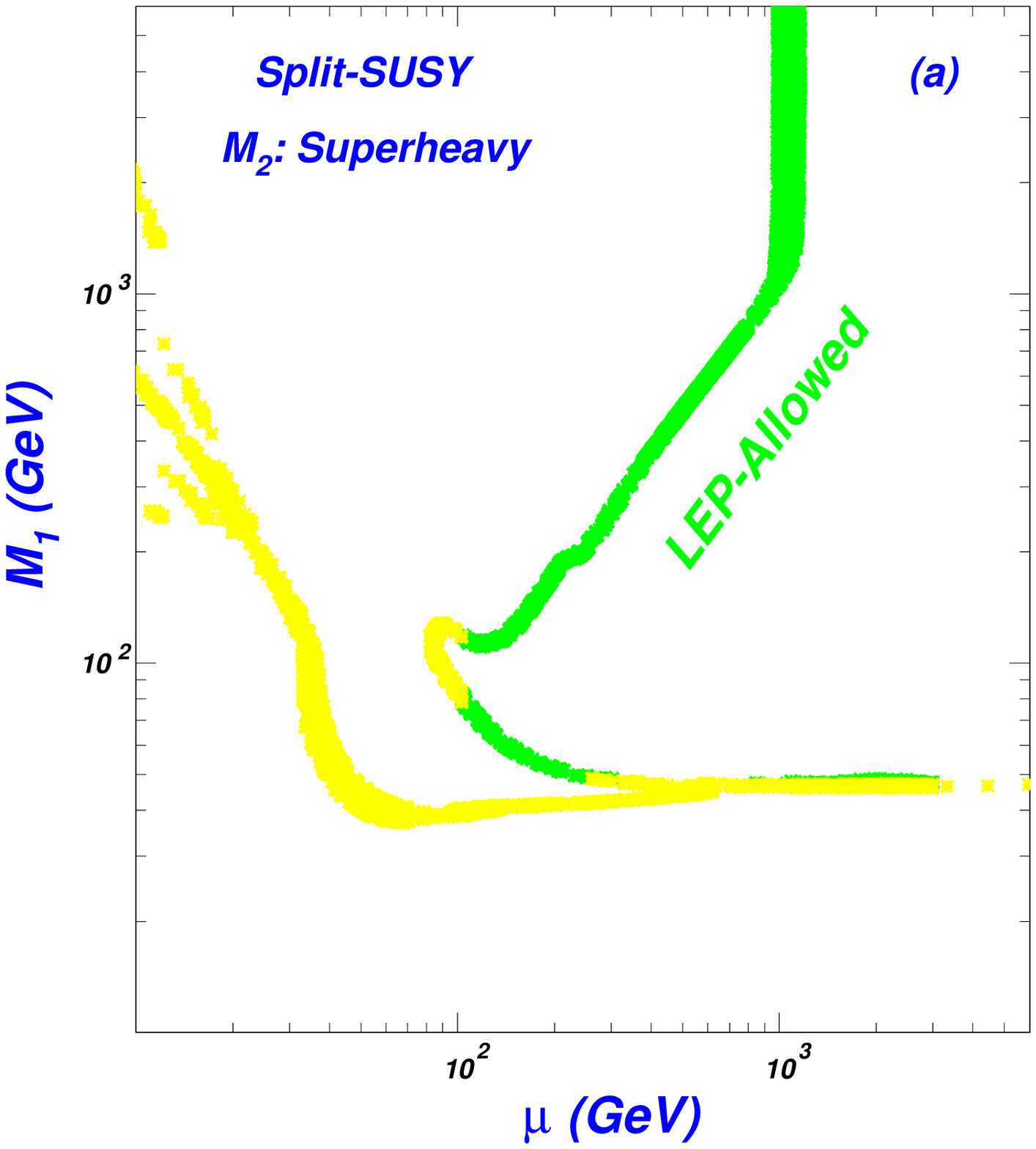,width=6.2cm, height=7.3cm}
                  \epsfig{file=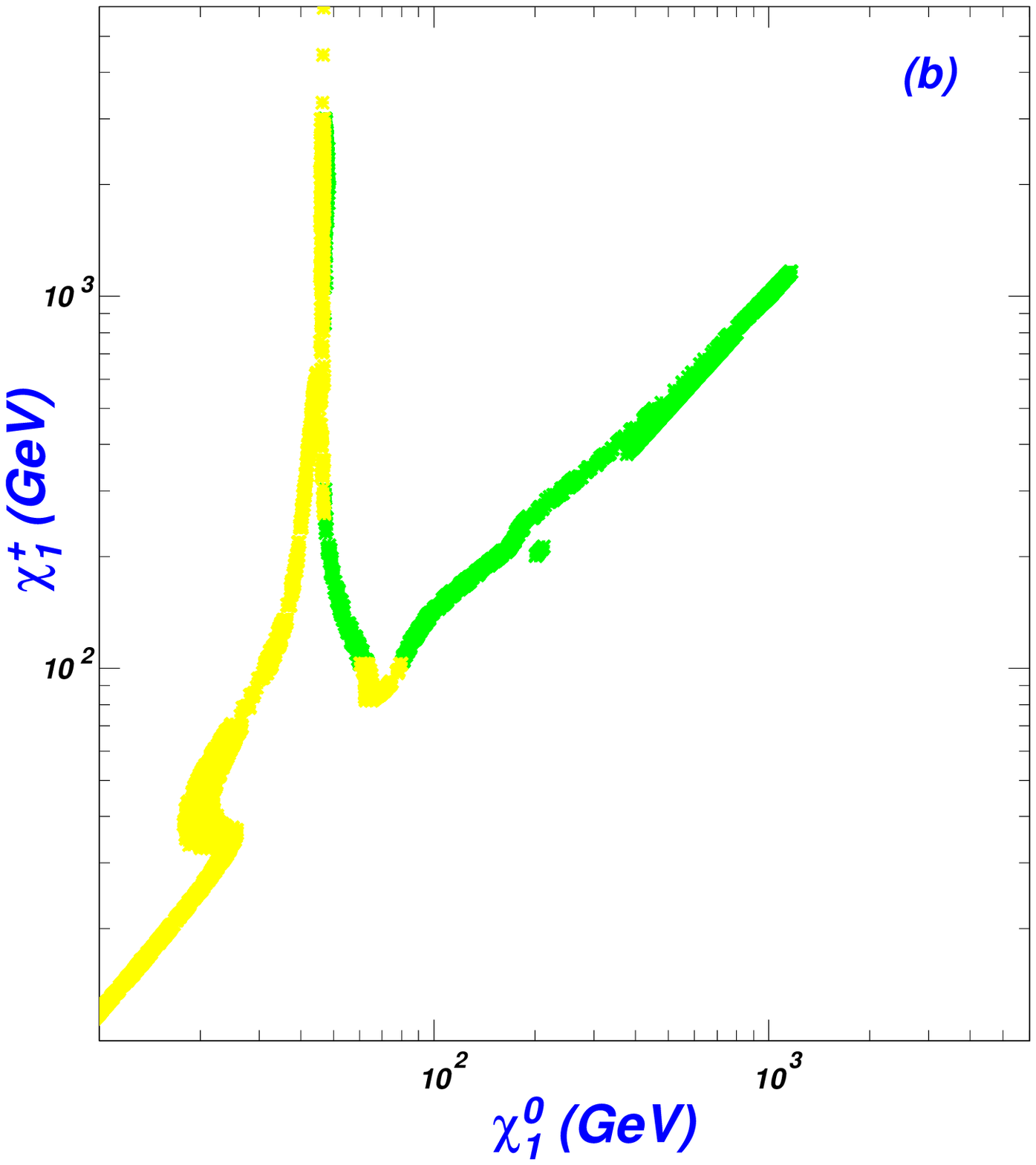,width=6.2cm, height=7.3cm}
                  \epsfig{file=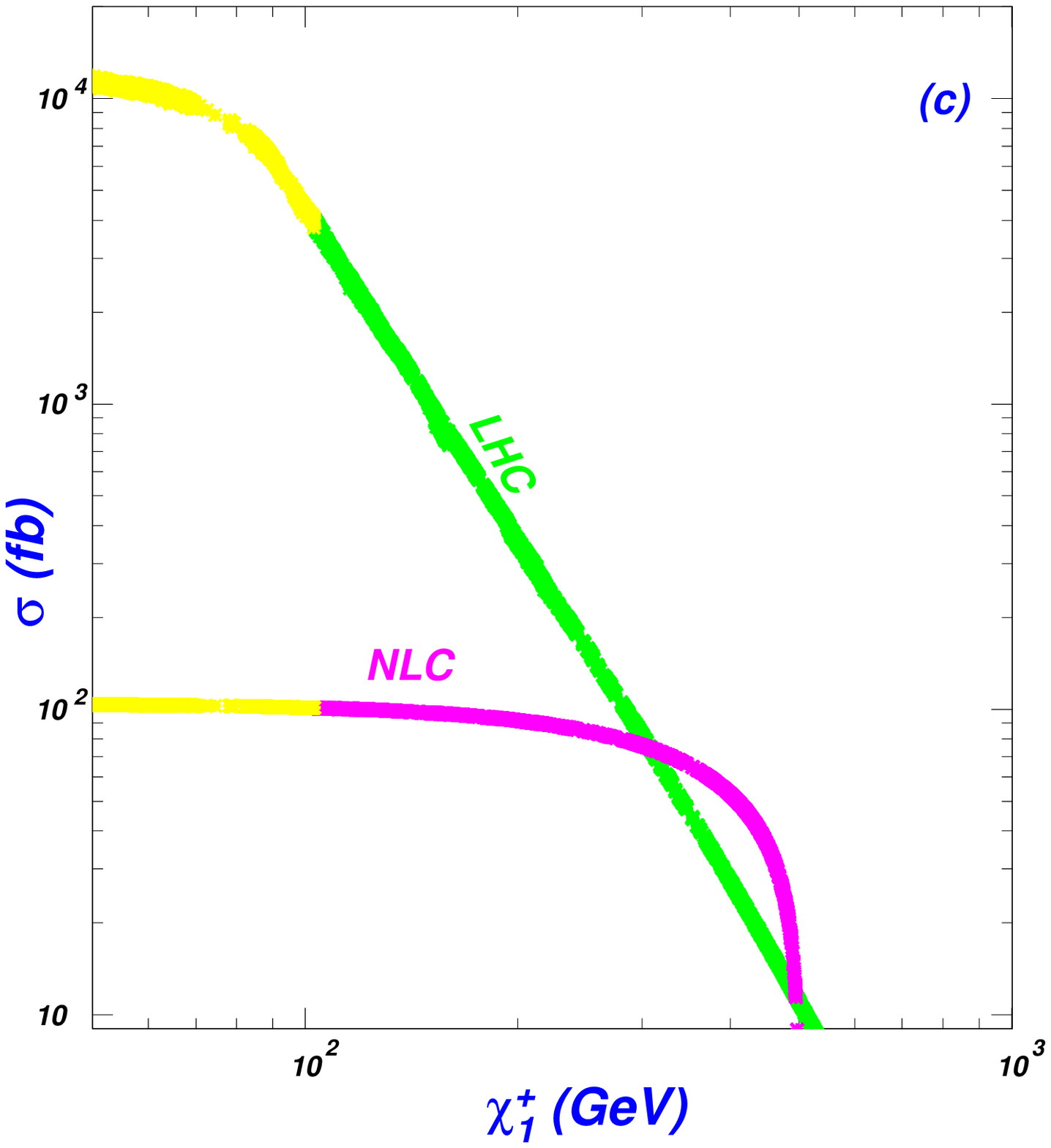,width=6.2cm, height=7.3cm}
 \vspace*{0.5cm}
\caption{The WMAP $2\sigma$ allowed region  (shaded area) in case of superheavy $M_2$.
         The light shaded region (yellow) is not allowed by LEP experiment.}
\end{figure}
\begin{figure}[tb]
 \hspace*{-1.5cm} \epsfig{file=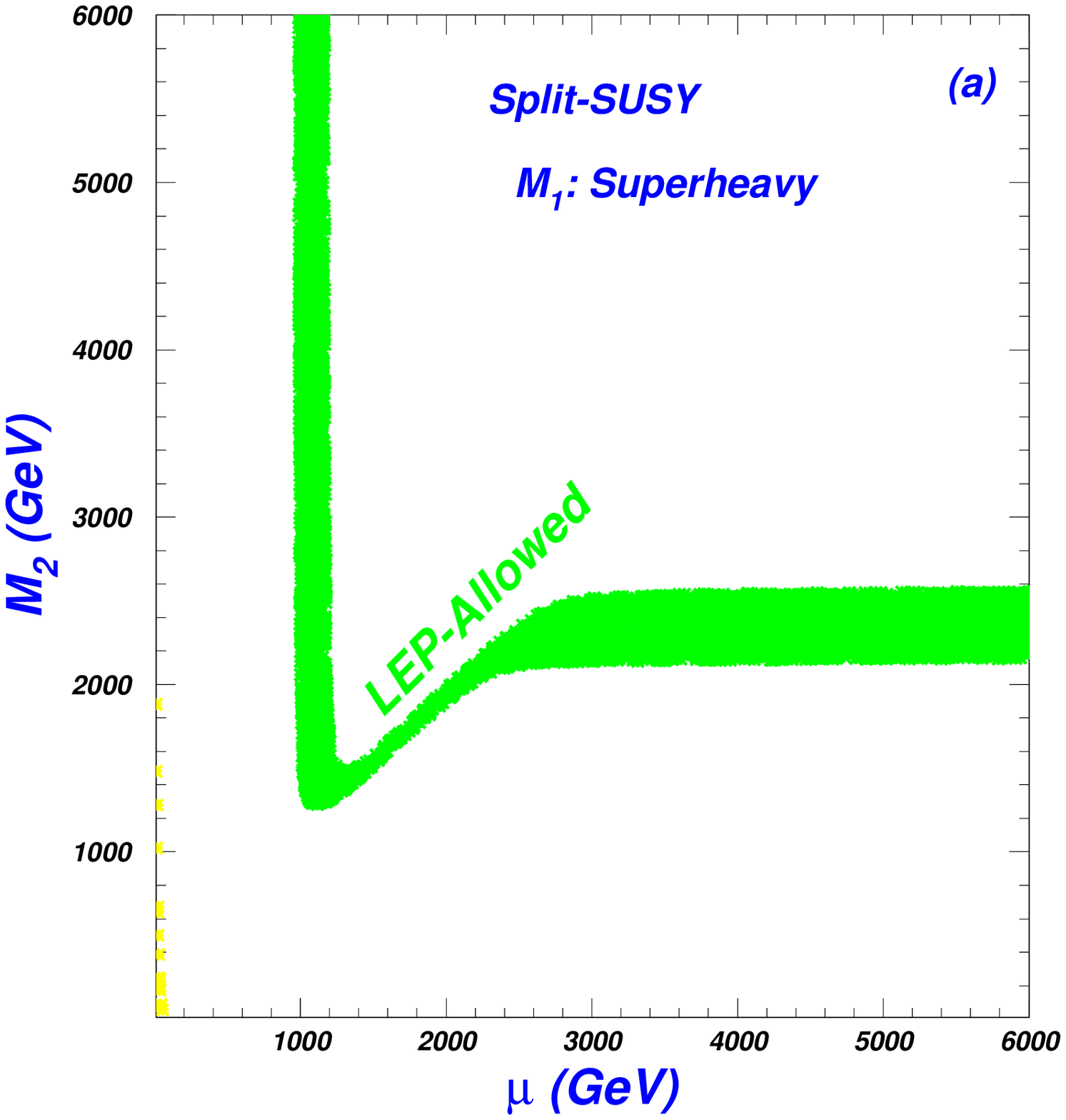,width=6.2cm, height=7.3cm}
                  \epsfig{file=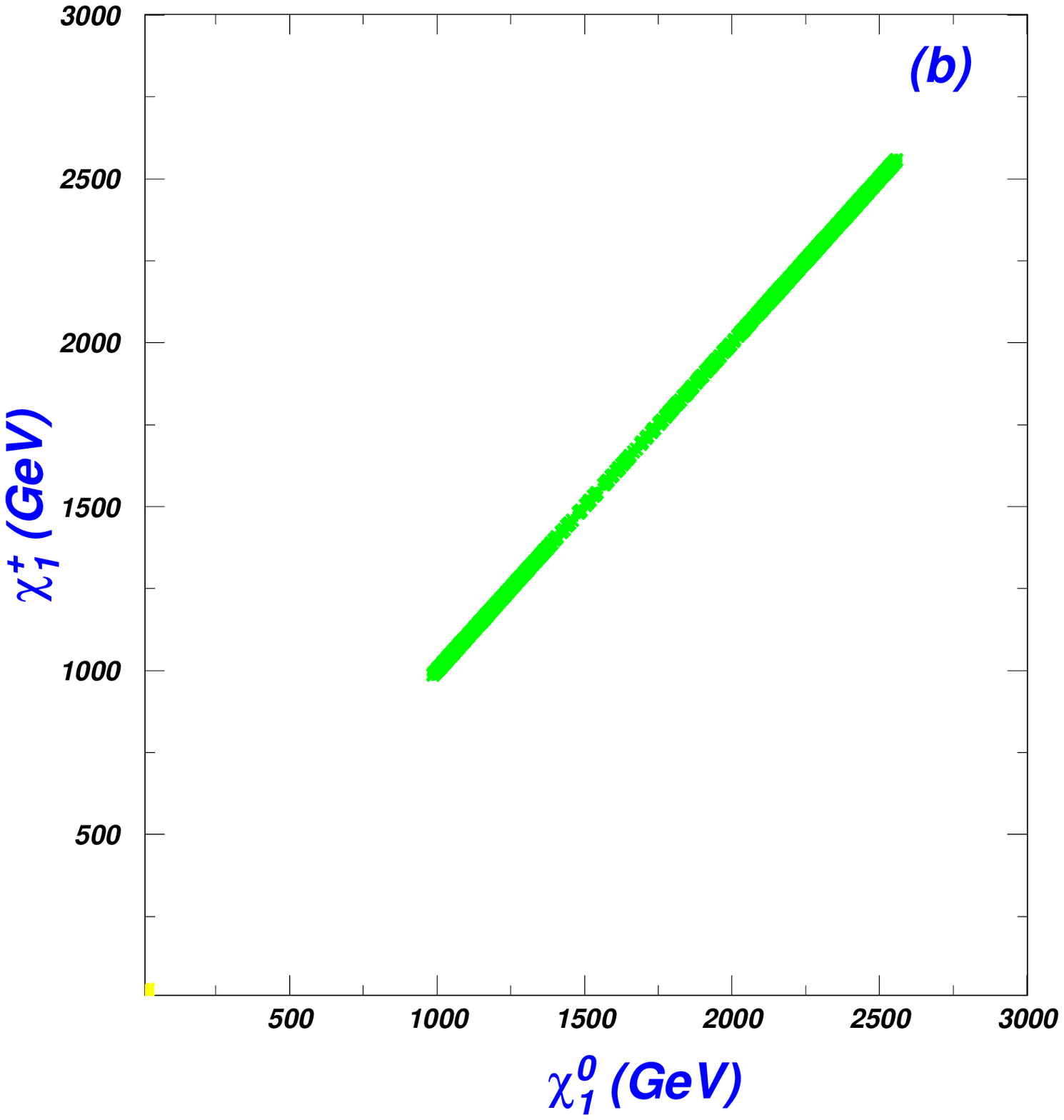,width=6.2cm, height=7.3cm}
                  \epsfig{file=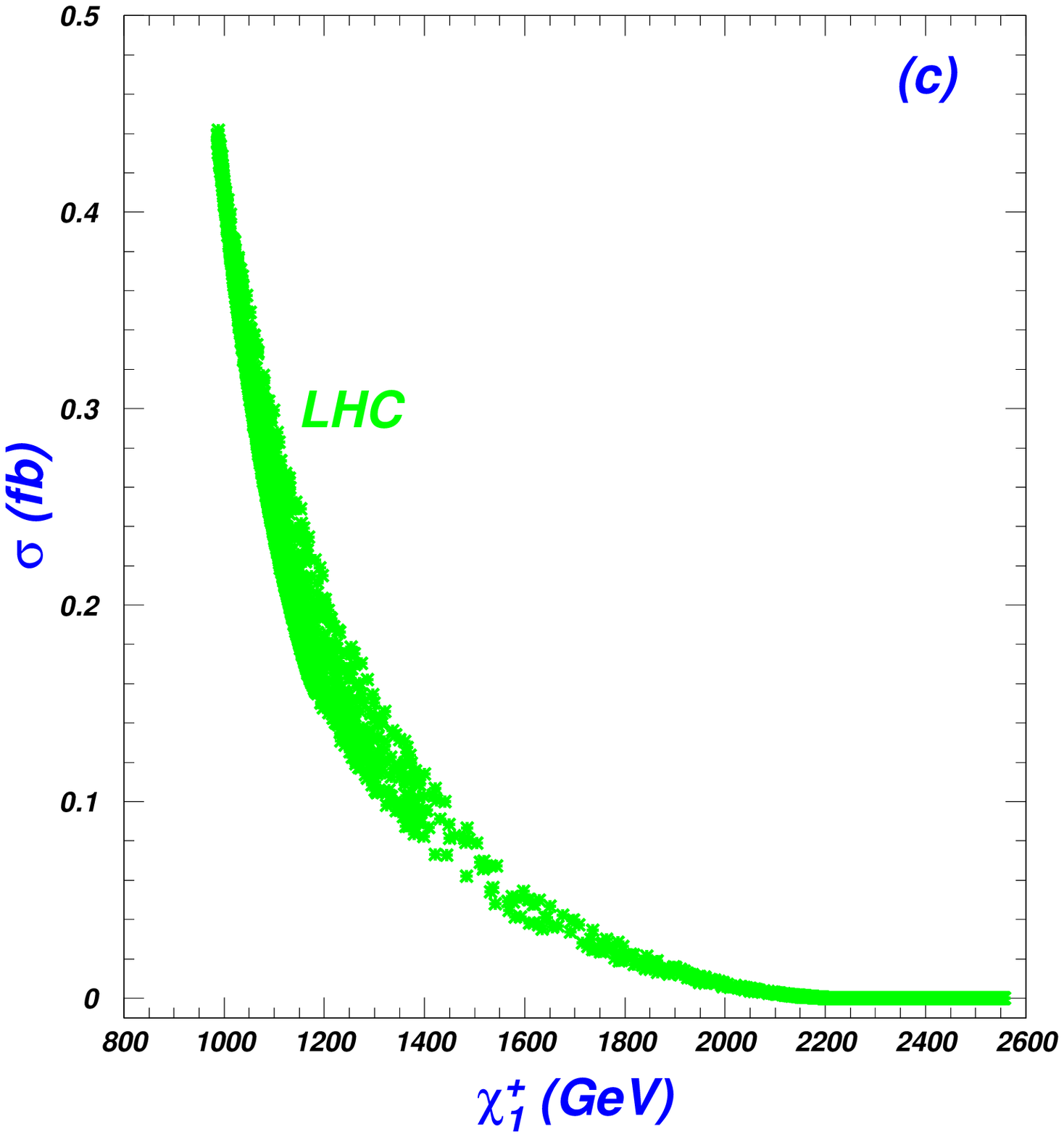,width=6.2cm, height=7.3cm}
 \vspace*{0.5cm}
\caption{ The WMAP $2\sigma$ allowed region  (shaded area) in case of superheavy $M_1$.
 The light shaded region (yellow) is not allowed by LEP experiment.}
\end{figure}

Note that while the chargino pair production with a rate of 100 fb at ILC
may not be hard to observe due to the clean enviroment of the ILC ( chargino pair 
production is regarded as a good way to test split supersymmetry at ILC \cite{zhu}),   
searching for the chargino pair production with a cross section of pb level
at the LHC may be quite challanging. The chargino $\chi^+_1$ decays into a neutralino $\chi^0_1$
and a pair of fermions (two jets or a charged lepton plus a neutrino). So the 
signature of chargino pair production is (i) two energetic leptons plus missing energy,
or (ii) one energetic lepton plus two jets plus missing energy. Let us take the latter
signature, i.e., $\ell+2j+P_T^{\rm miss}$, as an example. The huge background comes from
$Wjj$. In order to substantially reduce this background, we may apply a cut
on the transverse mass defined by
\begin{eqnarray}
m_T = \sqrt{ (P_T^\ell+P_T^{\rm miss})^2
- (\vec P_T^\ell+\vec P_T^{\rm miss})^2} .
\end{eqnarray}
$m_T$ is always less than $M_W$ (and peaks just below $M_W$) if the only missing 
energy comes from a neutrino from $W$ decay,
which is the case for the $Wjj$ background events.
For the signal, $m_T$ is spread about equally above and below $M_W$, due to the 
large extra missing energy from the neutralinos. Therefore, we may, for example, 
require $m_T > 90$ GeV. Given the importance of chargino pair production as
a test of split supersymmetry at the LHC, detailed Monte Carlo studies with
the consideration of various backgrounds are needed, which is beyond the scope 
of this work.        
  
\section{Conclusion}
In split supersymmetry, gauginos and Higgsinos are the only supersymetric 
particles which are possibly accessible at the LHC or the ILC collider.
The masses of these particles are subject to the stringent constraints from
the cosmic dark matter. Under the assumption that the lightst neutralino is the LSP 
and constitute the dark matter in the universe, we scrutinized the dark matter 
constraints on the masses of charginos and neutralinos. 
We considered several cases:
(1) $M_1=M_2/2$; (2) $\mu$ is superheavy; (3) $M_2$ is superheavy; and 
(4) $M_1$ is superheavy.
We found that the lightest chargino $\chi^+_1$ must be lighter than about 1 TeV 
in the first case and about 2 or 3 TeV in other cases. 
In the first three cases, the corresponding production rate of the chargino pair 
at the LHC (ILC) can reach the level of pb (100 fb) in some parts of the allowed region
and thus hopefully observable. But in the last case, the chargino must be heavier 
than about 1 TeV and thus has a too small production rate to be observale at the LHC.
So, overall, there is no guarantee to find the charginos of split supersymmetry 
at the LHC or ILC collider.  

\section*{Acknowlegement}
We thank Bing-Lin Young and Fuqiang Xu for discussions. 
This work is supported in part by National Natural Science Foundation of China.


\begin{thebibliography}{11}
\vspace*{-1.3cm}
\bibitem{split} N. Arkani-Hamed, S. Dimopoulos, hep-th/0405159;
                G.F. Giudice, A. Romanino, \NPB699, 65 (2004).
                N. Arkani-Hamed, S. Dimopoulos, G. F. Giudice, A. Romanino, \NPB709, 3 (2005).
\bibitem{split-natural}  C. Kokorelis, hep-th/0406258;  \NPB732, 341 (2006);
                        B. Kors, P. Nath, \NPB711, 112 (2005);
                        K. S. Babu, T. Enkhbat, B. Mukhopadhyaya, \NPB720, 47 (2005);
                        E. Dudas, S. K. Vempati, \NPB727, 139 (2005).
\bibitem{senatore} L. Senatore, \PRD71, 103510  (2005);                   
                   A. Masiero, S. Profumo, P. Ullio, \NPB712, 86 (2005).
\bibitem{wang} F. Wang, W. Y. Wang, J. M. Yang, hep-ph/0507172.
\bibitem{pierce} A. Pierce, \PRD70,  075006 (2004);
                 A. Arvanitaki, P. W. Graham, hep-ph/0411376.
\bibitem{profumo}  A. Masiero, S. Profumo, P. Ullio, \NPB712, 86 (2005).
\bibitem{gunion}  H. E. Haber and G. L. Kane, Phys. Rep. {\bf 117}, 75 (1985);
                  J. F. Gunion and H. E. Haber, \NPB272, 1 (1986).
\bibitem{pdg} S. Eidelman, {\it et al.}, Particle Data Group, \PLB592, 1 (2004).
\bibitem{lep2-web} LEP2 SUSY Working Group, http://lepsusy.web.cern.ch/lepsusy/.
\bibitem{cao} J. Cao, J. M. Yang, \PRD71, 111701 (2005).
\bibitem{feng} J. L. Feng, A. Rajaraman, F. Takayama, \PRL91, 011302 (2003); \PRD68, 063504 (2003);
               J. L. Feng, S. Su, F. Takayama, hep-ph/0404198; hep-ph/0404231;
               F. Wang, J. M. Yang, \EPJC38, 129 (2004);
               K. Hamaguchi, Y. Kuno, T. Nakaya, M. M. Nojiri, \PRD70, 115007  (2004). 
\bibitem{darksusy} P. Gondolo, J. Edsjo, L. Bergstrom, P. Ullio, M. Schelke, E. A. Baltz,
                    astro-ph/0406204.
\bibitem{micro} G. Belanger, F. Boudjema, A. Pukhov, A. Semenov, hep-ph/0112278.
\bibitem{wmap}  WMAP Collaboration, Astrophys. J. Suppl. {\bf 148}, 1 (2003); 
                                         {\bf 148}, 175 (2003).
\bibitem{char-lhc}
  W.~Beenakker, R.~H\"opker, M.~Spira and P.~M.~Zerwas, \NPB492, 51 (1997);
  W.~Beenakker, M.~Klasen, M.~Kr\"amer, T.~Plehn, M.~Spira and P.~M.~Zerwas, \PRL83, 3780 (1999);  
       Prospino2.0, http://pheno.physics.wisc.edu/$\sim$plehn
\bibitem{ILC} K.~Abe {\it et al.},  ACFA Linear Collider Working Group, hep-ph/0109166;
              T.~Abe {\it et al.},  American Linear Collider Working Group, hep-ex/0106056;
              J.~A.~Aguilar-Saavedra {\it et al.},  ECFA/DESY LC Physics Working Group,
                                                    hep-ph/0106315.
\bibitem{split-split} K. Cheung, C.-W. Chiang, \PRD71, 095003 (2005).
\bibitem{mu} J.~E.~Kim and H.~P.~Nilles, \PLB138, 150 (1984);
             Y.~Nir, \PLB354, 107 (1995);
             M.~Cvetic and P.~Langacker, \PRD54, 3570 (1996)
\bibitem{zhu} S. H. Zhu, \PLB604, 207 (2004);
              W.~Kilian, T.~Plehn, P.~Richardson and E.~Schmidt, \EPJC39, 229 (2005).
\end{thebibliography}
\end{document}